\documentclass[pra,aps,a4paper]{revtex4}

\begin{document}

\date{26th June, 2008 }

\newcommand\pkl{\tilde{\phi }_k^L}
\newcommand\pkr{\tilde{\phi }_k^R}
\newcommand\pkm{\tilde{\phi }_k^-}
\newcommand\pkp{\tilde{\phi }_k^+}
\newcommand\domegak{\delta \omega _k^-}
\newcommand\doa{\delta \omega _k^A}

\title{Dephasing in two decoupled one-dimensional Bose-Einstein
condensates and the subexponential decay of the interwell coherence}
\author{I.E. Mazets$^{1,2}$ and   J. Schmiedmayer$^{1}$}

\affiliation{ $^1$ Atominstitut der \"Osterreichischen
Universit\"aten, TU Wien, A--1020 Vienna, Austria \\
$^2$  A.F. Ioffe Physico-Technical Institute, 194021
St. Petersburg, Russia}

\begin{abstract}
{We provide a simple physical picture of the loss of coherence
between two coherently split one-dimensional Bose-Einstein
condensates. The source of the dephasing is identified with
nonlinear corrections to the elementary excitation energies in
either of the two independent condensates. We retrieve the result
by Burkov, Lukin and Demler [{Phys. Rev. Lett.} {\bf 98}, 
200404 (2007)] on the subexponential 
decay of the cocherence $\propto \exp
[-(t/t_0)^{2/3}]$ for the large time $t$, however, the scaling of
 $t_0$ differs.\\
 \vspace*{3mm} 
condensates; collective and hydrodynamic
excitations, superfluid flow.  }
\end{abstract}

\maketitle

Effectively one-dimensional systems of ultra cold atoms are a
model systems to study the fundamental processes of the coherent
dynamics and (de) coherence in interacting many body systems. In
addition in the limit of zero temperature they are a primary
example of the exactly integrable Lieb-Liniger model \cite{L1,L2}.

Recently experimental progress on both, optical lattices
\cite{OLreview} and atom chips \cite{ACRev,FZ}, allow to confine
ultra cold atoms in strongly elongated  traps with $\omega _r \gg
\omega_z$ ($\omega _r$, $\omega _z$ being the frequencies of the
radial and longitudinal confinement, respectively).  These traps
are an ideal system for studying 1D physics as long as both the
temperature $T$ and chemical potential $\mu$ are small compared to
the energy scale given by the transverse confinement: $ \mu <
\hbar \omega_r, \; k_BT<\hbar \omega_r $.  Optical lattices enable
study of global properties of ensembles of 1d systems down to very
small atom numbers and into the strongly correlated regime, atom
chips allow to study the properties and dynamics of single 1d
systems.

Strong inhibition of thermalization, a signature of integrability,
was observed with bosons deep in the 1D regime \cite{dw1}, and
interference experiments on atom chips with pairs of weakly
interacting Bose gases, easily fulfilling the above conditions for
one dimensionality, allowed to study the dynamics of (de)
coherence \cite{H1} and the interplay between thermal and quantum
noise \cite{H2}.

The special interest in the decoherence in such effectively
one-dimensional ultracold atomic systems is rooted in the fact
that they display decoherence and thermalization despite being at
the first glance a prime example of the exactly integrable
Lieb-Liniger model \cite{L1,L2}. In a recent paper
\cite{recentwe}, we have shown that even if the temperature and
chemical potential are well below the energy of the radial
excitation ($ \mu < \hbar \omega_r, \; k_BT<\hbar \omega_r $)
radial modes can be excited virtually.  These virtually excited
radial modes give rise to effective three-body velocity-changing
collisions which lead to thermalization and the break down of
integrability. The typical thermalization scale for typical 1d
atom chip experiments \cite{H1,H2,va} is in the order of 100 ms
even when thermalization due to two body collisions is completely
suppressed. However the dephasing dynamics observed in such
coherently split ultracold 1d atomic clouds \cite{H1,MIT} is even
one order of magnitude faster .

In addition, the experiment on the time evolution of interference
between two coherently split one-dimensional (1D) atomic
Bose-Einstein condensates (BECs) \cite{H1} revealed an surprising
sub-exponential decay of the inter-well coherence $\langle
\hat{\psi }_R^\dag \hat{\psi }_L \rangle$.
\begin{equation}
\langle \hat{\psi }_R^\dag (x,t)\hat{\psi }_L(x,t) \rangle \propto
\exp [-(t/t_0)^{\alpha}] . \label{eq:1}
\end{equation}
The measured decay exponent $\alpha \approx 2/3$ is in agreement with
the theoretical calculations of Burkov {\em et al.} \cite{BLD}
which predicts $\alpha = 2/3$.  In their theoretical approach the
inter-well coherence decay is treated in the terms of the heat
flow between symmetric, $\hat{\psi }_{+}=[\hat{\psi
}_{L}+\hat{\psi }_{R}]/\sqrt{2}$, and antisymmetric, $\hat{\psi
}_{-}=[\hat{\psi }_{L}+\hat{\psi }_{R}]/\sqrt{2}$, modes. The two
individual (fully split) condensates are designated as right (R)
and left (R), $\hat{\psi }_{R,L}(x,t)$ being the corresponding
atomic field annihilation operator in the coordinate
representation.

In the present work we choose a different, more intuitive way of
describing the system and consider {\em dephasing} of two
independent integrable systems with inter-correlated initial
conditions. As in Burkov {\em et al.} \cite{BLD} we consider the
weak interaction limit, which, together with the finite size of
the system, allows for a finite condensed fraction, in contrast to
the Tonks limit or the case of an infinitely long quasicondensate.
In what follows we use the system of units where Planck's and
Boltzmann's constants are set to 1.

After the splitting, the system consists of two independent BECs,
each being described by the Hamiltonian
\begin{equation}
\hat{\cal H}=\int dx\, \left[ \frac 1{2m}
\left( \frac \partial {\partial x}\hat{\psi }_j ^\dag \right)
\left( \frac \partial {\partial x}\hat{\psi }_j \right) +
\frac {g_{1D}}2 \hat{\psi }_j ^\dag \hat{\psi }_j ^\dag \hat{\psi }_j
\hat{\psi }_j\right] , \qquad j=L,R,
\label{eq:2}
\end{equation}
where $m$ is the atomic mass, $g_{1D}$ is the 1D coupling constant.
The main contribution to the coherence reduction stems from
phase fluctuation, so that
\begin{equation}
\langle \hat{\psi }_R^\dag (x,t)\hat{\psi }_L(x,t) \rangle \approx
\bar{n}\exp \left( - \langle \hat{\phi }_-^2\rangle \right) ,
\label{eq:3}
\end{equation}
where the phase operators $\hat{\phi }_\pm =(\hat{\phi }_L\pm
\hat{\phi }_R)/\sqrt{2}$ and their conjugate density fluctuations
operators $\hat{ n} _\pm =(\hat{n }_L\pm \hat{n }_R)/\sqrt{2}$ are
defined via $\hat{\psi }_j=(\bar{n}+\hat{n }_j)^{1/2}\exp (i
\hat{\phi }_j)$, $j=L,R$, $\bar{n}$ being the average 1D BEC density.
In what follows, we consider only classical (thermal-like) fluctuations
and therefore omit operator notations, writing simply $\phi _j$ etc.

The local fluctuations can be expanded in plane waves,
\begin{equation}
\phi _\pm (x,t)=\frac 1{\sqrt{\cal L}}\sum _k \tilde{\phi }_k^\pm (t)
\exp (ikx),
\label{eq:4}
\end{equation}
${\cal L}$ being the quantization length. The linearization of Eq.
(\ref{eq:2}) yields Bogoliubov spectrum $\omega _k=\sqrt{k^2/(2m)[
k^2/(2m)+2mc^2]}$, $c$ being the speed of sound. In what follows
we consider BEC at temperatures below the chemical potential, so
that we assume phonon spectrum of excitations of the uniform 1D
Bose gas at rest,  $\omega _k=c|k|$.

The key idea of our treatment is to recall that the local density
and velocity fields in BEC fluctuate, thus giving rise to the
{\em random non-linear} corrections to the phonon frequency.
Phonons propagating in the left and right condensates interact
with different fluctuations, and this is the source of dephasing.

The phonon energy depends on the BEC local density $n=\bar{n}+
\delta n$ via the speed of
sound $c=\sqrt{g_{1D}{n}/m}$. Also the fluctuating local velocity
$V=m^{-1}\partial \phi /\partial x$ in the BEC contributes to
the random energy correction as the
advective term, so that the total correction to the
average phonon energy $c(\bar{n})|k|$ is
\begin{equation}
\delta \omega _k =\left[ \frac d{d\bar{n}}c(\bar{n})\delta n+V
\right] |k| =
c|k|\left( \frac {\delta  {n}}{2\bar{n}}+\frac Vc \right) .
\label{eq:5}
\end{equation}
Fluctuations of the density and local velocity differ for the
right and left BECs, because the splitting process is never
completely adiabatic (the degree of nonadiabaticity has been recently
quantified by Polkovnikov and Gritsev \cite{gri}),
and initial quantum (zero-point) fluctuations of $\phi ^-$ and $n^-$
can be amplified to the values, comparable at low momenta to
the initial thermal (classical) fluctuations of $\phi ^+$ and $n^+$.
Still quantum fluctuations on their own can initiate the
decay of the interwell coherence \cite{alt} (on a different time scale).
Since the density and local velocity fluctuations in
the right and left BECs are different,
the random energy shift is also different for the excitations
propagating in these two BECs:
\begin{equation}
\delta \omega _k^{L,R} =\delta \omega _k^{+} \pm \domegak  ,
\label{eq:6}
\end{equation}
Using Eq. (\ref{eq:4}), we obtain
\begin{equation}
\pkm (t)=\exp \left( -ic|k|t-i\int _0^tdt^\prime \,
\delta \omega _k^+ \right) \left[
\pkm (0) \cos J_k(t) - i\pkp (0)\sin J_k(t) \right] ,
\label{eq:7}
\end{equation}
where
\begin{equation}
J_k(t) =\int _0^tdt^\prime \, \domegak (t^\prime ) .
\label{eq:8}
\end{equation}
Assuming that fluctuations in $k$-mode at $t=0$ are not correlated
with the fluctuations of $\domegak $ at later times, we obtain
from Eq. (\ref{eq:7})
\begin{equation}
\langle [\pkm (t)]^2 \rangle -\langle [\pkm (0)]^2\rangle =
\left \{ \langle [\pkp (0)]^2 \rangle -\langle [\pkm (0)]^2\rangle
\right \} \left \langle \sin ^2J_k(t)\right \rangle .
\label{eq:9}
\end{equation}
In the short-time limit Eq. (\ref{eq:9}) reduces to
\begin{equation}
\frac \partial {\partial t} \langle (\pkm )^2 \rangle =
\left \{ \langle (\pkp )^2 \rangle -\langle (\pkm )^2\rangle
\right \} \frac 1t\left \langle J_k^2(t)\right \rangle .
\label{eq:10}
\end{equation}
Moreover, since the system under consideration is closed and
consists of two independent (L and R) integrable subsystems,
the sum $\langle (\pkl )^2\rangle +\langle (\pkr)^2 =
\langle (\pkp )^2 \rangle +\langle (\pkm )^2\rangle $ is
time-independent, and we obtain
\begin{equation}
\frac \partial {\partial t} \left[ \langle (\pkm )^2 \rangle -
\langle (\pkp )^2\rangle \right] = 2
\left \{ \langle (\pkm )^2 \rangle -\langle (\pkp )^2\rangle
\right \}  \frac 1t\left \langle J_k(t)^2\right \rangle .
\label{eq:11}
\end{equation}

Since the correlation function for the random interwell
frequency shift is
\begin{equation}
\langle \domegak (t^\prime )\domegak (t^{\prime \prime})\rangle
= (\doa ) ^2f_k(t^{\prime \prime}-t^{ \prime}),
\label{eq:12}
\end{equation}
where $\doa $ characterizes the amplitude of the frequency shift
fluctuations for the mode with given $k$, and $f_k(\tau )$
is the correlation function in dimensionless form, equal to 1
at $\tau =0$ and rapidly approaching 0 if its argument exceeds
certain correlation time $\tau _k^c$. Then Eq. (\ref{eq:11})
yields
\begin{eqnarray}
\langle (\pkm )^2 \rangle -\langle (\pkp )^2\rangle &\propto &
\exp (-\Gamma _kt^2/\tau _k^c), ~~t\ll \tau _k^c,
\label{eq:13} \\
\langle (\pkm )^2 \rangle -\langle (\pkp )^2\rangle &\propto &
\exp (-2\Gamma _kt), ~~~~t\, ^>_\sim \,  \tau _k^c ,
\label{eq:14}
\end{eqnarray}
where the dephasing rate is
\begin{equation}
\Gamma _k\approx (\doa ) ^2\tau _k ^c .
\label{eq:15}
\end{equation}

Assuming that the condensate splitting is close to adiabatic and,
hence,
\begin{equation}
\langle [\pkm (0)]^2 \rangle \ll \langle [\pkp (0)]^2\rangle
\label{eq:16}
\end{equation}
and neglecting the change (approximately by a factor of 2) of
$\langle (\pkp )^2 \rangle $ in the course of the system evolution
(similar approximations are assumed in Ref. \cite{BLD}), we obtain
in the limit $t\, ^>_\sim \, \tau _k^c$
\begin{equation}
\langle (\pkm )^2 \rangle =\langle (\pkp )^2\rangle \left[ 1-
\exp (-2\Gamma _k t)\right] .
\label{eq:17}
\end{equation}

The remaining question is to determine $\doa $ and $\tau _k^c$.
To estimate $\doa $ at least roughly, we apply the following
method. Instead of plain waves, we consider a wave packet
centered at the momentum $k$ and having the momentum
uncertainty $\Delta k\sim k$. Such a width allows us to choose
the shape of the wave packet close to the minimum-uncertainty
wave packet, that allows us to localize it on a spatial scale
of the order of $2\pi /k$. Then we can identify the
random energy shift experienced by such a wave packet with that
of the phonon with momentum $k$.

In such a context, obviously, only the density and velocity
fluctuations at wavelengths longer than $2\pi /k$ contribute
to $\domegak $. Fluctuations at shorter wavelengths are
effectively averaged out and cause no influence to the
dynamics of a wave packet of a spatial extension $\sim 2\pi /k$.
Taking into account that fluctuations at
different momenta are not correlated and relacing sum over
discrete state by integration over continuous spectrum, we obtain
\begin{equation}
\doa \,^2=c^2k^2\int _{-k}^k \frac {dk^\prime }{2\pi }\,
\left[ \frac {\langle (\tilde{n}_{k^\prime }^-)^2
\rangle }{8\bar{n}^2}+
\frac{k^2\langle (\tilde{\phi }_{k^\prime }^- )^2
\rangle }{2m^2c^2}\right] .
\label{eq:18}
\end{equation}
Here $\tilde{n}_k^-$ is the Fourier component of the density
fluctuations introduced analogously to Eq. (\ref{eq:4}).

The remaining issue is to estimate the correlation time $\tau
_k^c$. Say, the wavepacket propagates along $x$ in the positive
direction. Half of the surrounding fluctuations propagates in the
opposite direction, bringing about a short correlation time scale
$\sim 1/(ck)$. However, half of the fluctuations co-propagate with
the wave packet, in the first approximation at the same velocity.
If there were no dephasing (with respect to each other) of
fluctuations at different momenta, the corresponding correlation
time would be infinite. However, longer-wavelength correlations
dephase as well, at a rate similar to that of the wave packet
under consideration. Averaging the dephasing rates over the
ensemble of fluctuations restricted to $|k^\prime |<|k|$, we
obtain the correlation time of the fluctuations affecting the
dynamics of phonons with momentum $k$ to be
\begin{equation}
\tau _k^c\approx \Gamma _k^{-1}.
\label{eq:19}
\end{equation}
The correlation time cannot be longer than given by Eq. (\ref{eq:19}),
because otherwise Eq. (\ref{eq:13}) holds instead of Eq. (\ref{eq:14}),
and the dephasing is slowed down significantly. Eqs.
(\ref{eq:15},~\ref{eq:19}) result in
\begin{equation}
\Gamma _k\approx \doa .
\label{eq:20}
\end{equation}
It is unlikely that deeper insight to the short-time dynamics
of nonlierly interacting modes of a 1D BEC can enhance the
dephasing rate further compared to Eq. (\ref{eq:20}), because
short-time dephasing given by Eq. (\ref{eq:13}) is of the form
$\exp (-\mathrm{const} \, t^2)\approx 1-\mathrm{const} \, t^2$,
and fast perturbation can cause only slowdown of the evolution
(Quantum Zeno effect), but not speed up (anti-Zeno effect) \cite{KK}.
Eqs. (\ref{eq:17},~\ref{eq:20}) yield finally the following
dephasing dynamics:
\begin{equation}
\langle (\pkm )^2 \rangle =\langle (\pkp )^2\rangle \left[ 1-
\exp (-\doa  t)\right]
\label{eq:21}
\end{equation}
(the factor of 2 in the decrement is omitted, because of
inexact nature of the estimations involved).

In the asymptotically long time limit we may estimate fluctuation
amplitudes from the 1D Bogoliubov treatment for the phononic
ensemble at temperature $T$ as illustrated in \cite{BT1D} (the
same assumption is taken in Ref. \cite{BLD}):
\begin{equation}
\frac 1{\bar{n}^2}\langle (\tilde{n}_{k}^-)^2\rangle =
\left( \frac k{mc}\right) ^2 \langle (\pkm )^2 \rangle \approx
\frac T{\mu \bar{n}},
\label{eq:22}
\end{equation}
where $\mu =mc^2$ is the zero-temperature approximation for the
chemical potenial. Substituting Eq. (\ref{eq:22}) into
Eq. (\ref{eq:18}), we obtain
\begin{equation}
\doa \approx \sqrt{ \frac T{2\pi \mu \bar{n} } }c^2|k|^{3/2} .
\label{eq:23}
\end{equation}
Finally, Eq. (\ref{eq:21}) takes the form
\begin{equation}
\langle (\pkm )^2 \rangle =\frac {mT}{\bar{n}k^2}\left[ 1-
\exp \left(-0.4 \, \sqrt{ \frac T{\mu \bar{n}} }
c|k|^{3/2} t\right) \right] ,
\label{eq:24}
\end{equation}
which is to a certain extent similar to Eq. (14) of Ref.
\cite{BLD}: in both cases the dephasing rate is proportional to
$|k|^{3/2}$, but there is also an important difference: the
relaxation rate predicted in Ref. \cite{BLD} is by the factor of
$\bar{n}/(mc)$ faster than our estimate. Eq. (\ref{eq:24})
has a generic form
\begin{equation}
\langle (\pkm )^2 \rangle =\frac {b_1}{k^2} {\cal F}(b_2|k|^{3/2}t),
\label{eq:24bis}
\end{equation}
where $b_1$, $b_2$ are certain constants and ${\cal F}(\Theta )$ is
a function that is finite at $\Theta \rightarrow \infty $ and
decreases faster than $\Theta ^(2/3)$ if $\Theta \rightarrow 0$,
to provide the convergence of the integral
\begin{equation}
\langle {\phi }_-^2\rangle  =
\int _{-k_{max} }^{k_{max} } \frac {dk}{2\pi }\,
\langle (\pkm )^2\rangle
\label{eq:3bis}
\end{equation}
that appears in the coherence factor Eq. (\ref{eq:3}). The
cut-off momentum $k_{max}$ depends on the temperature and
chemical potential and is of the order of $mc$ if $T\sim \mu $.
For $t\gg T^{-1}, \, \mu ^{-1}$ we can substitute
$k_{max}$ by $\infty $.
Changing the integration variable from momentum $k$ to the
dimensionless time $\Theta = b_2|k|^{3/2}t$ we obtain
\begin{equation}
\langle {\phi }_-^2\rangle  =
(b_2t)^{2/3}\frac {2b_1}{3\pi } \int _0^\infty d\Theta \,
\Theta ^{-5/3} {\cal F}(\Theta ) \equiv (t/t_0)^{2/3},
\label{eq:27}
\end{equation}
that provides the subexponential decay Eq. (\ref{eq:1}) with
$\alpha =2/3$. Comparing Eqs.
(\ref{eq:24}) and (\ref{eq:24})  we find the
scaling time $t_0$ of the subexponential decay Eq. (\ref{eq:1}):
\begin{equation}
t_0\approx 3.2\, \frac \mu {T^2}\left( \frac {\bar{n}}{mc}\right)
^2= 3.2\, \frac {\bar{n}^2}{mT^2} . \label{eq:25}
\end{equation}
Note that the time limit  $t\gg T^{-1}, \, \mu ^{-1}$ is not
sufficient for Eq. (\ref{eq:27}) to hold, since the
latter is derived under assumption of significant thermalization
of the antisymmetric mode that should happen at $t\sim t_0$.
Our estimation of $t_0$ is by the factor of $0.24\,
\bar{n}/(mc)>1$ longer than that of Ref. \cite{BLD}  and the
scaling with experimental parameters are also different. Recalling
that $\mu =g_{1D}\bar{n}$ with $g_{1D}\propto \sqrt{\nu _\perp }$,
$\nu _\perp $ being the radial trapping frequency of the atomic
waveguide and regarding it, the 1D number density and
experimentally obtained $t_0$ as input parameters, we obtain
$T\propto \bar{n}/t_0$, whereas Ref. \cite{BLD} gives $T\sim
\bar{n}^{3/2}\sqrt{\nu _\perp}/t_0$.

Let us now compare our results to the findings of Burkov, Lukin
and Demler \cite{BLD} in more detail:

In the theoretical approach of Ref. \cite{BLD}, the inter-well
coherence decay is treated in the terms of the heat flow between
symmetric, $\hat{\psi }_{+}=[\hat{\psi }_{L}+\hat{\psi
}_{R}]/\sqrt{2}$, and antisymmetric, $\hat{\psi }_{-}=[\hat{\psi
}_{L}+\hat{\psi }_{R}]/\sqrt{2}$, modes. This point of view is
counterintuitive for two completely split one-dimensional BECs
which are two independent {close to \em exactly integrable}
systems \cite{L1,L2}.  The thermalization times calculated, even
when including the new virtual 3-body collisions \cite{recentwe},
are much longer then the observed decoherence. The termalization
time scale should be the one at which the heat flow arguments
should not apply.

The results of Ref. \cite{BLD} seem to imply some unphysical
consequences: If we consider $T\sim \mu $ and estimate the
dephasing rate for phonons with the energy close to the chemical
potential  Ref. \cite{BLD} gives a rate in the order of
$\sqrt{\bar{n}/(mc)}\mu \gg \mu $ (since the experimentally
accessible 1D BECs are characterized by $\bar{n}/(mc)\approx 30$),
which is quite counterintuitive: the phonons with $k\sim mc$
become over damped, and their dephasing rate exceeds any frequency
scale available in the system with $T\sim \mu $. Under the same
condition our theory predicts dephasing rate $\sim \mu
/\sqrt{\bar{n}/(mc)} \ll \mu $.

The time scale $t_0$ as suggested by our estimates implies the
reconsideration of the temperature estimations for the
experimental data of Ref. \cite{H1}. Our model suggests that the
actual final temperatures were higher by a factor of order 2 than
it was concluded from \cite{BLD}. This may easily  happen, since
the mechanisms of the heating of a BEC during the splitting (which
is adaibatic only partially, taking into account its time scale
$\sim 10$ ms) are not yet explored and understood.

This difference between the prediction of the timescales and the
scaling with experimental parameters between our model and the
calculations by Burkov, Lukin and Demler \cite{BLD} demands both
new more detailed experiments over a wider parameter range and an
comprehensive numerical calculation of the splitting and coherence
dynamics.  New ways to measure temperature using the statistics of
interference patterns \cite{H2}, and more refined RF potentials
for atom manipulation \cite{H0} will greatly extend the capability
for experimental investigations.

This work is supported by the MIDAS STREP and the FWF (Lise
Meitner fellowship for I.E.M.). We thank T. Schumm and H.-P.
Stimming for fruitful discussions.

\end{document}